\documentclass{aastex}

\slugcomment{To appear in ApJ}

  \shorttitle{Implications of the Optical Observations of Isolated Neutron Stars}
  \shortauthors{Shearer \& Golden}

\begin{document}
\title{Implications of the Optical Observations of Isolated Neutron Stars}
 \author{Andy  Shearer, Aaron Golden}
 \email{andy.shearer@nuigalway.ie, aaron.golden@nuigalway.ie}
\affil{The National University of Ireland, Galway, Newcastle Road, Galway, Ireland}

\begin{abstract}

We show that observations of pulsars with pulsed optical emission
indicate that the peak flux scales according to the magnetic field
strength at the light cylinder. The derived relationships indicate
that the emission mechanism is common across all of the observed
pulsars with periods ranging from 33ms to 385 ms and ages of
1000-300,000 years. It is noted that similar trends exist for $\gamma$
ray pulsars. Furthermore the model proposed by Pacini (1971) and
developed by Pacini and Salvati (1983,1987) still has validity and
gives an adequate explanation of the optical phenomena.

\end{abstract}

\section{Introduction}

Since the first optical observations of the Crab pulsar in the late
1960s (\cite{coc69}), 7 more pulsars have been observed in the optical
bandpass. Of these only 4 have been seen to pulsate optically
(PSR0540-69, \cite{mid85}; Vela, \cite{wall77}; PSR 0656+14,
\cite{shear97}; Geminga, \cite{shear98}).  Whilst the three remaining
pulsars (PSR0950+08, PSR 1929+10 and PSR1055-52) are thought to be
mainly thermal emitters in the optical regime, the pulsating objects
are either purely magnetospheric emitters or a mixture of thermal and
non-thermal sources. Two of these five pulsars (PSR0540-69 and the
Crab) are certainly too distant to have any detectable optical thermal
emission using currently available technologies. For those with
distances less than a kiloparsec (Geminga, Vela and PSR0656+14) only
the Vela pulsar is too bright for its emission to be anything other
than non-thermal in origin. For the remaining two pulsars (PSR0656+14
and Geminga) their optical emission is thought to be a mix of thermal
and non-thermal fluxes (\cite{gold99} and \cite{mar98}). However for
the faintest of the two, Geminga, there is disagreement about the
level of thermal emission.  Broad-band photometric studies have been
interpreted as being consistent with the Rayleigh-Jeans tail combined
with a cyclotron emission feature (\cite{mig98}).  These authors
suggest that the optical emission is predominantly thermal with an
embedded ion cyclotron resonance feature at about 5500 \AA. More
recently, \cite{jac99}, derive a complex phenomenological model for
this feature coming from a thin plasma above the pulsar's polar
cap. Their derived surface magnetic field using this model is
consistent with the independently determined 'spin-down' field ($B
\propto (P\dot{P})^{0.5}$). Conversely, spectroscopic studies have
indicated that the emission is a mix of thermal and non-thermal
radiation with the non-thermal dominating in the optical region
(\cite{mar98}). These authors do not see the proposed emission
feature, but, there is a statistically weak absorption feature at
6400\AA. Temporal studies of the pulsars indicate that for the Crab
and Vela pulsars the peak of $\gamma$-ray and optical pulses are
coincident. A similar coincidence is noted for the Geminga pulsar
although the optical signal is of lower significance. This can
interpreted as indicating a common source location for both $\gamma$
and optical photons. Recent studies of the plateau of the Crab main
pulse in the optical indicate that it is of limited ($<15$ kms)
extent (\cite{gold00}).

Many suggestions have been made concerning the non-thermal optical
emission process for these young and middle-aged pulsars. Despite many
years of detailed theoretical studies and more recently limited
numerical simulations, no convincing models have been derived which
explain all of the high energy properties. There are similar problems
in the radio but as the emission mechanism is radically different
(being coherent) only the high energy emission will be considered
here. 

In recent years a number of groups have carried out detailed
simulations of the various high-energy emission processes. These
models divide into two broad groups - those with acceleration and
emission low in the magnetosphere (polar cap models) and those with
the acceleration nearer to the light cylinder (outer-gap models). Both
models have problems explaining the observed features of the limited
selection of high energy emitters. Both models also suffer from
arbitrary assumptions in terms of the sustainability of the outer-gap
and the orientation of the pulsar's magnetic field to both the
observers line of sight and the rotation axis. Furthermore some
observational evidence, see for example \cite{eik97}, severely limits
the applicability of the outer-gap to the emission from the Crab
pulsar. However these models have their successes - the total
polar-cap emission can be understood in terms of the Goldreich-Julian
current ({\cite{gold69}) from in or around the cap; the Crab
polarisation sweep is accurately reproduced by an outer-gap variant of
\cite{rom95}. However, the most successful model which adequately
explains most of the high-energy phenomena (both in terms of its
elegance and longevity) has been proposed by \cite{pac71} and in
modified form by Pacini and Salvati (1983 PS83 \& 1987 PS87
hereafter). In general they proposed that the high energy emission
comes from relativistic electrons radiating via synchrotron processes
in the outer regions of the magnetosphere and that the luminosity is a
strong function of the period ($\propto P^{-10}$ in the original
formulation). In this paper we examine the validity of their approach
and show that it still adequately explains the observed phenomena.

It is the failure of the detailed models to explain the high energy
emission that has prompted this work. We have taken a phenomenological
approach to test whether Pacini type scaling is still applicable. Our
approach has been to try to restrict the effects of geometry by taking
the peak luminosity as a scaling parameter rather than the total
luminosity. In this regard we are removing the duty cycle term from
PS87. It is our opinion that to first order the peak emission
represents the local power density along the observer's line of site
and hence reflects more accurately emission processes within a
pulsar's magnetosphere. Previous work in this area (see for example
\cite{gold95}) looked at the total efficiency, spectral index and
found no reasonable correlation with standard pulsar parameters - age,
period and spin down rate. Their work was hampered by not including
geometry and being restricted to the then three known pulsed optical
emitters. Since then the number of pulsars with observed pulsed
optical emission has increased to five.

\section{The Phenomenology of Magnetospheric Emission}

The three optically brightest pulsars (Crab, Vela and PSR 0540-69) are
also amongst the youngest. However all these pulsars have very
different pulse shapes resulting in a very different ratio between the
integrated flux and the peak flux. In this work we will use the peak
emission as the primary flux parameter. A number of definitions can
exist for this and in this context we have taken the 95\%-95\% level
of the primary pulse. To first order, this correlates well with the
luminosity per pulse divided by the Full-Width at Half-Maximum
(FWHM). For the Crab pulsar the cusp is effectively flat over this
region (\cite{gold00a}). The FWHM can be considered to scale with the
pulsar duty cycle. Our proposition is that the peak flux represents
the local power density within the emitting region with minimal
effects from geometrical considerations such as observer line of sight
and magnetic and rotational axis orientation. Table 1 shows the basic
parameters for these objects including the peak emission (taken as the
emission in the 95\% - 95\% portion of the largest peak). Their
distances imply that the thermal emission should be low, in all cases
$<$ 1\% of the observed emission, and any such contribution is
negligible.
       
Of all the optical pulsars Geminga is perhaps the most
controversial. Early observations (\cite{hal88}, and \cite{big88})
indicated that Geminga was an $\approx$ 25.5 m$_V$ object.  Subsequent
observations including HST photometry appeared to support a thermal
origin for the optical photons, albeit requiring a cyclotron resonance
feature in the optical (\cite{mig98}).  The high-speed optical
observations of \cite{shear98} combined with spectroscopic
observations (\cite{mar98}) contrast with this view. Figures 1 and 2
show how this apparent contradiction could have arisen. Figure 1,
based upon data from \cite{mig98} shows the integrated photometry. It
is possible to fit a black body Rayleigh-Jeans tail through this, but
only with the fitting of a cyclotron resonance emission feature at
about 5500 \AA. Figure 2 however shows the same points plotted on top
of the Martin et al spectra, where we have also included the pulsed B
point (\cite{shear98}). All the data points are consistent within
their error bars. This combined data set indicates a fairly steep
spectrum (with spectral index of 1.9) consistent with magnetospheric
emission and a weak thermal component, without the requirement for a
cyclotron feature. It was on the basis of these results that
\cite{gold99} were able to give an upper limit of R$_\infty$ of about
10km, by considering the upper limits to the unpulsed fraction of the
optical emission from Geminga as an upper limit for the thermal
emission.  A simpler view is that the soft X-ray and EUV data is
predominantly thermal emission with a magnetospheric component
becoming dominant at about 3500 \AA. Figure 3 illustrates this with
the thermal and magnetospheric components from \cite{gold99} combined
with the broad-band points of \cite{mig98} and \cite{shear98}. We
should also note that the optical pulses seen by \cite{shear98} are
coincident with the observed $\gamma$ ray peaks. Clearly more
spectroscopic data, particularly if temporally resolved, will be
crucial to determining the exact mix of thermal and non-thermal
emission.

As regards PSR 0656+14 the pulsar is generally agreed to be
predominately a non-thermal emitter in the optical, becoming
predominantly a thermal emitter at wavelengths shorter than about
3000\AA  (\cite{pav97}).  However, there is a discrepancy between the
radio distance based upon the dispersion measure and the best fits to
the X-ray data. From radio dispersion measure a distance of $ 760 \pm
190 pc $ can be derived, at odds with the X-ray distance of $250-280
pc$ from $N_H$ galactic models. Clearly more observations are needed
to determine a parallax - both radio and optical. This discrepancy in
distance leads to an ambiguity in the total luminosity. For this paper
we have taken the lower distance measure.

In Table 1 we have indicated the peak flux normalised to the Crab
pulsar. When comparing the individual pulsars, which will have a range
of viewing angles and differing magnetic/rotation axes, we have to
account for the viewing angle (related to the pulse duty cycle and
separation) as well as the total flux. Furthermore for each source, we
have to account for the shape of the pulsar's spectrum. For all the
observed pulsars, with the exception of Vela, the low energy cut-off is
above 7000 \AA (\cite{nas97}). PS87 indicates for emission above the
low energy cut-off that the ratio of the fluxes from two different
pulsars can be given by Equation 1 if one ignores the effect of duty
cycle and pitch angle (the suffices refer to each pulsar).  Here
$F_{\nu,n}$ refers to the flux at the observed frequency $\nu$ for
pulsar n.  Similarly for the magnetic field strength, B, and period,
P. The observed energy spectrum exponent is given by $\alpha_n$. The
duty cycle can be accounted for by only considering the peak
emission. The pitch angle is beyond the scope of this work and
assumed to first order to be invariant. Equation 2 shows the the same
formulation for the outer field case.  \\

\begin{equation}
           {{F_{\nu,2}} \over {F_{\nu,1}}} \propto ({\nu_{1,0} \over {\nu}})^{\alpha_2-\alpha_1} 
({{B_{2,0}} \over {B_{1,0}}})^{4-\alpha _2} ({{P_{2}} \over {P_{1}}})^{3\alpha-9} \\ 
\end{equation}

Scaling to the transverse field would give:-

\begin{equation}
           {{F_{\nu,2}} \over {F_{\nu,1}}} \propto ({\nu_{1,0} \over {\nu}})^{\alpha_2-\alpha_1} 
({{B_{2,0}} \over {B_{1,0}}})^{4-\alpha _2} ({{P_{2}} \over {P_{1}}})^{3} \\ 
\end{equation}

Given the observed peak luminosities we investigated the correlation
between the peak emission and the surface field and the tangential light
cylinder field. Figure 4 shows the relationship between the peak
luminosity and the outer magnetic field, $B_T$, Goldreich-Julian
current and canonical age (${P} \over {2\dot{P}}$). A clear
correlation is seen with all these parameters. In this paper we
investigate the implications of a correlation between the peak
luminosity and the transverse field. We accept that the strong
correlation with G-J current would underpin both emission from polar cap as
well as outer regions.

A weighted regression of the form :-

Peak Luminosity $\propto B_T^{\beta}$ 

was performed for the empirical peak luminosity leading to a
relationship of the form :- 

Peak Luminosity $\propto B_T^{2.86 \pm 0.12 }$ 

which is significant at the 99.5\% level. Figure 5 shows the predicted
peak luminosity from Equation 2 against our observed peak luminosity
accounting for the differing observed energy spectrum exponent at 4500
\AA.  The slope is 0.95 $\pm$ 0.04 and significant at the 99 \% level.
Importantly, we note that the flattening of the peak luminosity
relationship with the outer field strength for the older, slower
pulsars, see Figure 4, is consistent with their having a significantly
steeper energy spectrum than the younger pulsars, see Table 1. Whilst,
from PS87, it is possible that this indicates that the emission zone is
optically thick at these frequencies, alternatively it might reflect
a larger emission region for these pulsars compared to the younger
ones. We also note that the generation parameter concept
(\cite{wei97}) indicates that the $\gamma$-ray spectral index scales
with the average number of generations within the $e^+$/$e^-$
cascade. Our observed correlation between peak luminosity and age
(which linearly scales with the generation parameter) and similarly
between spectral index and age indicates a link between optical
luminosity and $\gamma$-ray luminosity phenomena.

We note as well as that similar trends can be seen in $\gamma$-rays.
Figure 6 shows the correlation between the peak $\gamma$ emission as a
function of transverse field. This indicates a regression of the form

Peak $\gamma$ Luminosity $\propto B_T^{0.86 \pm 0.17}$ \\ 
\\
significant at the 99.3 \% level, consistent with the observed steeper
distribution seen in high energy $\gamma$-rays - see Table 2 and Equation 2.

It seems clear from both polarisation studies (\cite{smith88};
\cite{rom95}) and from this work that we expect that the optical emission
zone is sited towards the outer magnetosphere. Timing studies of the
size of the Crab pulse plateau indicates a restricted emission volume
($\approx$ 15 kms in lateral extent) (\cite{gold00}). Importantly,
the simple relationships we have derived here indicate that there is
no need to invoke complex models for this high energy
emission. Observed variations in spectral index, pulse shape and
polarisation can be understood in terms of geometrical and absorption
factors rather than differences in the production mechanism.

Combining our results and those of \cite{gold95}, we can begin to
understand how the high energy emission process ages. Goldoni et al
compared the known spectral indices and efficiencies in both the
optical and $\gamma$-ray regions. They noted that the spectral index
flattened with age for the $\gamma$-ray pulsars whilst the reverse was
true for the optically emitting systems. They also noted a similar
trend reversal for the efficiency, with the $\gamma$-ray pulsars
becoming more efficient with age. We note (see the bottom panel in
Figures 4 and 6) a similar behaviour with the peak emission. From the
temporal coincidence between the $\gamma$-ray and optical pulses, it
seems likely that the source location is similar for both
mechanisms. One explanation is that we have a position where
by from the same electron population there are two emission processes
- expected if we have curvature for the $\gamma$-ray photons and
synchrotron for the optical ones. It seems likely that the optical
photon spectrum has been further modified to produce the reversal in
spectral index with age. The region over which the scattering can
occur would scale with the size of the magnetosphere and hence with
age. With the outer magnetosphere fields for these pulsars being $<
10^6 $ G, electron cyclotron scattering is not an option. However
synchrotron self-absorbtion could explain the observed features. In
essence we would expect the most marked flattening to be for the Crab
pulsar, where the outer field strengths are of order $10^6$ G, and less
so for the slower and older systems.

These results (both optical and $\gamma$-rays) are consistent with a
model where the $\gamma$ and optical emission is coming from the last
open-field line at some constant fraction of the light cylinder. The
drop in efficiency with age for the production of optical photons
points towards an absorbing process in the outer magnetosphere.
Clearly more optical and $\gamma$-ray observations are needed to
confirm these trends.

\section{Conclusion}

We have shown that the peak optical luminosity is the key important
parameter that scales with other observed pulsar attributes.  We note
that similar behaviour can be seen in the $\gamma$-ray regime.  To
first order we confirm that the model first proposed by Pacini in 1971
still has validity. There is a proviso that there is a strong
correlation with the Goldreich-Julian current which underpins both
Pacini scaling and other models. From a more detailed anaysis of the
EGRET database \cite{mcl00} showed a similar functional form to our
derived relationship for the total luminosity against period and
surface magnetic field. We also show in Table 3 what the expected
luminosity would be from a number of X and $\gamma$-ray emitting
pulsars. We have chosen these on the basis of their known duty cycles
in the high energy regime. We have also considered the expected flux
from the anomalous X-ray pulsars, a class of magnetars. If the
observed $P$ and $\dot{P}$ relationship can be interpreted in same way
as for normal pulsars to derive a canonical surface field then we can
estimate what the expected luminosity of these object would be. The
derived luminosity is very low - a reflection of the weak light
cylinder field. A recent VLT observation (\cite{hull0}) has found
no optical candidates down to 25.5 R magnitude - consistent with no
magnetospheric emission and predominantly thermal output. Problems
remain however as to the source of the thermal emission. We note that,
in the outer magnetosphere of these objects, the magnetic field is far
too weak for optical non-thermal emission processes.

Of crucial importance in the future will be the determination of the
low-energy cutoff and the polarisation sweep through the optical
pulse.  Also of interest will be the shape of the pulse - in
particular the size of any plateau which scales as the size of the
emission zone. Optical observations of millisecond pulsars will also
be important, to see if they follow the age or field trends noted
here. Of interest in this regard is the result from X-ray observations
which indicate that the luminosity of millisecond pulsars scales in a
similar fashion to normal pulsars (\cite{bec97}). All of these
parameters are measurable with existing and emerging technologies and
telescopes.  

Finally, the recent detection of a 16 ms pulsar in the LMC (PSR
J0537-6910 (\cite{mars98}) which has defied optical detection despite
its low period can possibly be understood in terms of its age
($\approx$ 5,000 years), which on the basis of Figure 3 indicates peak
luminosity a factor of $\approx 10^5$ times lower than the Crab. This is
at the limit of some of the recent optical searches - \cite{mig00} and
\cite{gou00}.

\acknowledgements

This work was supported by the Enterprise Ireland Basic Grant scheme
whose assistance is gratefully acknowledged. Ray Butler is thanked for
his helped during the preparation of the manuscript. He and Padraig
O'Connor are thanked for useful comments during the course of this
work.  The anonymous referee is thanked for comments which improved an
early draft of this paper.

\begin{figure*}
\plotone{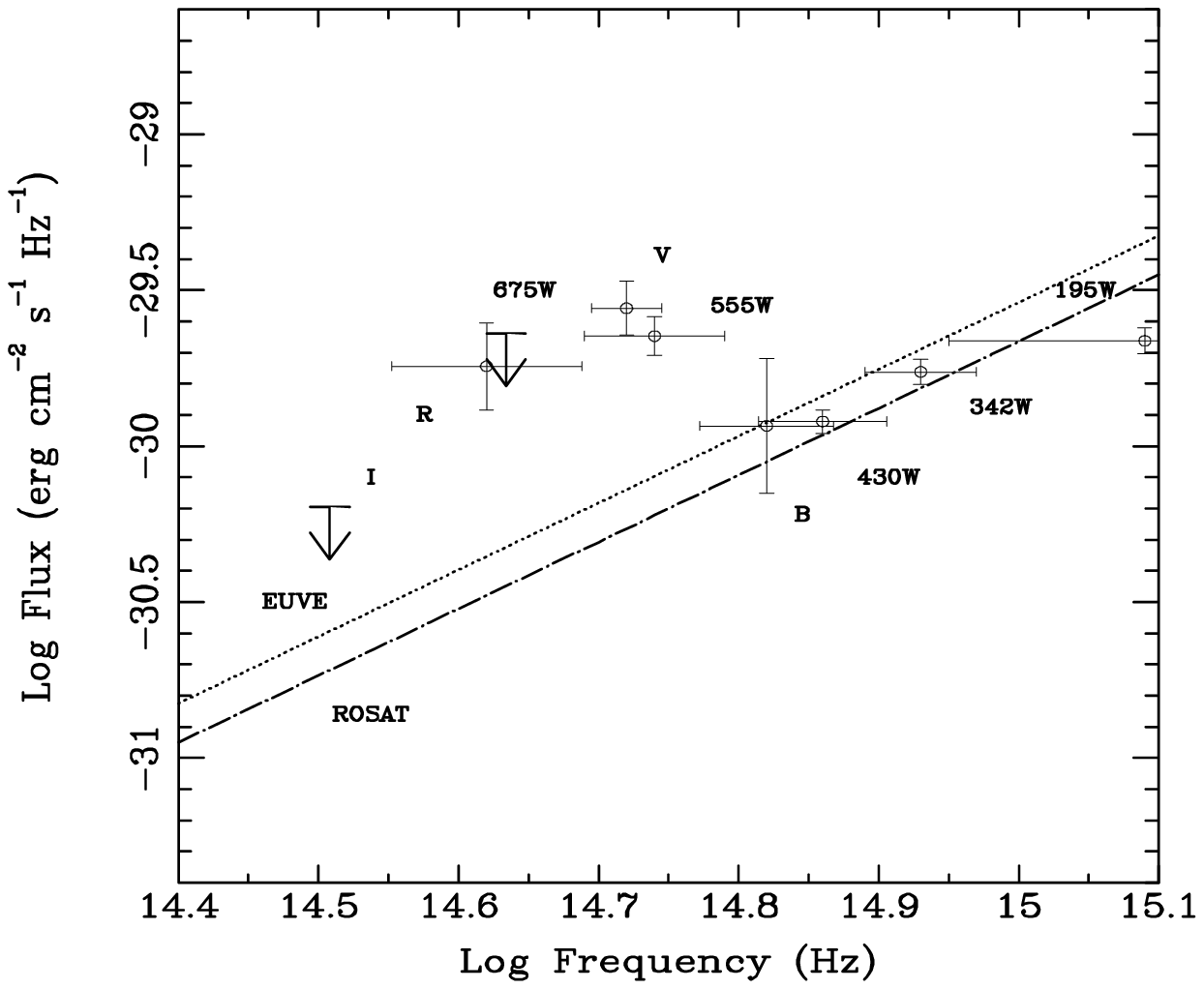}
\caption{Photometry of Geminga. The integrated photometry and the
thermal fit to the ROSAT X-Ray and EUVE data are both shown. }
\end{figure*}

\begin{figure*} 
\plotone{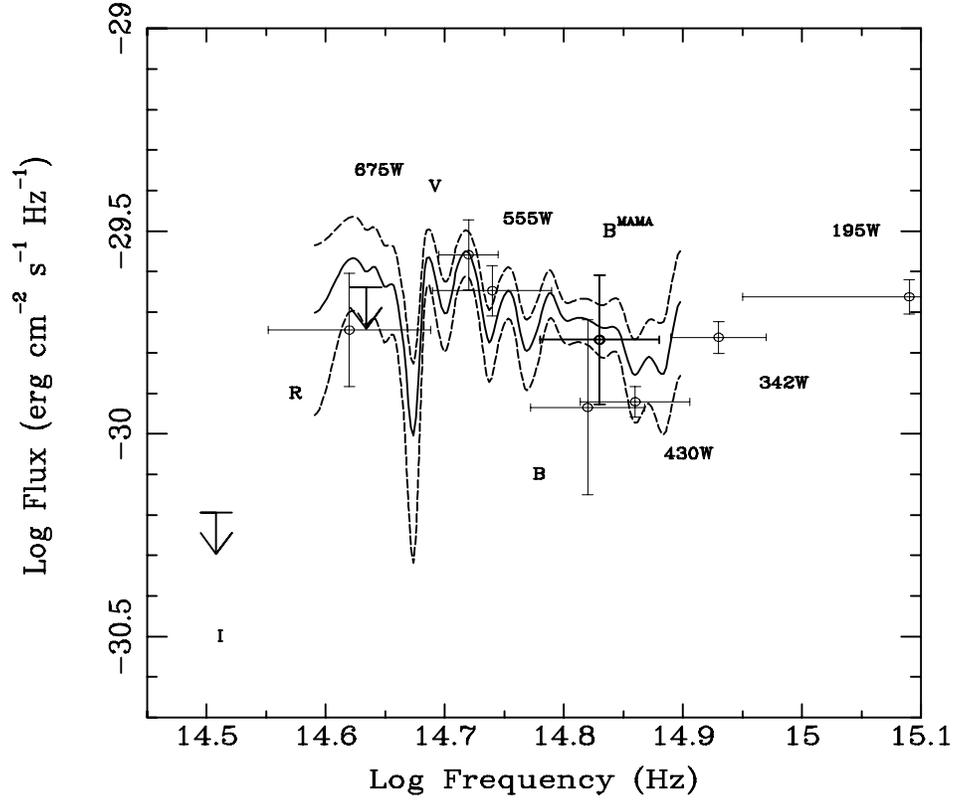} 
\caption{Spectra and Photometry of
Geminga.  The spectrum taken from \cite{mar98} (solid line) and 1
$\sigma$ error limits (dotted line) are shown. Note the agreement
between it and the integrated photometry from Figure 1. $B^{MAMA}$ is
the pulsed flux from \cite{shear98} and other points are from
\cite{mig98} and \cite{big96}.} 
\end{figure*}

\begin{figure*} 
\plotone{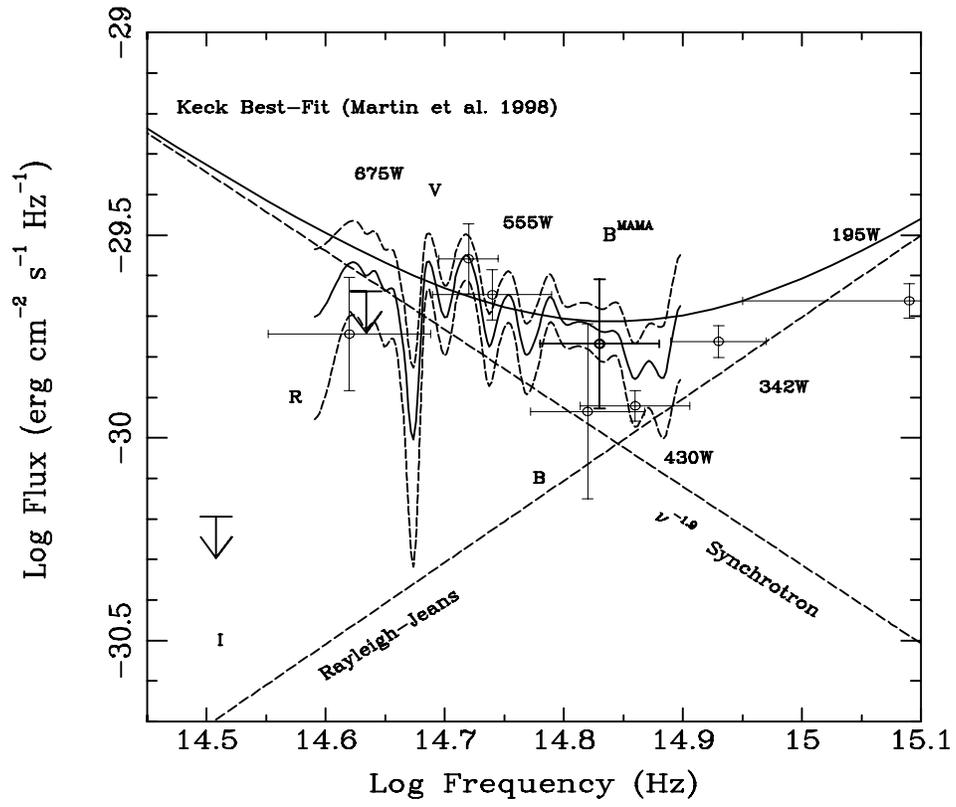} 
\caption{Derived thermal 
and non-thermal components for Geminga. Also plotted are the
broad-band points points from figures 1 and 2.}  
\end{figure*}

\begin{figure} \plotone{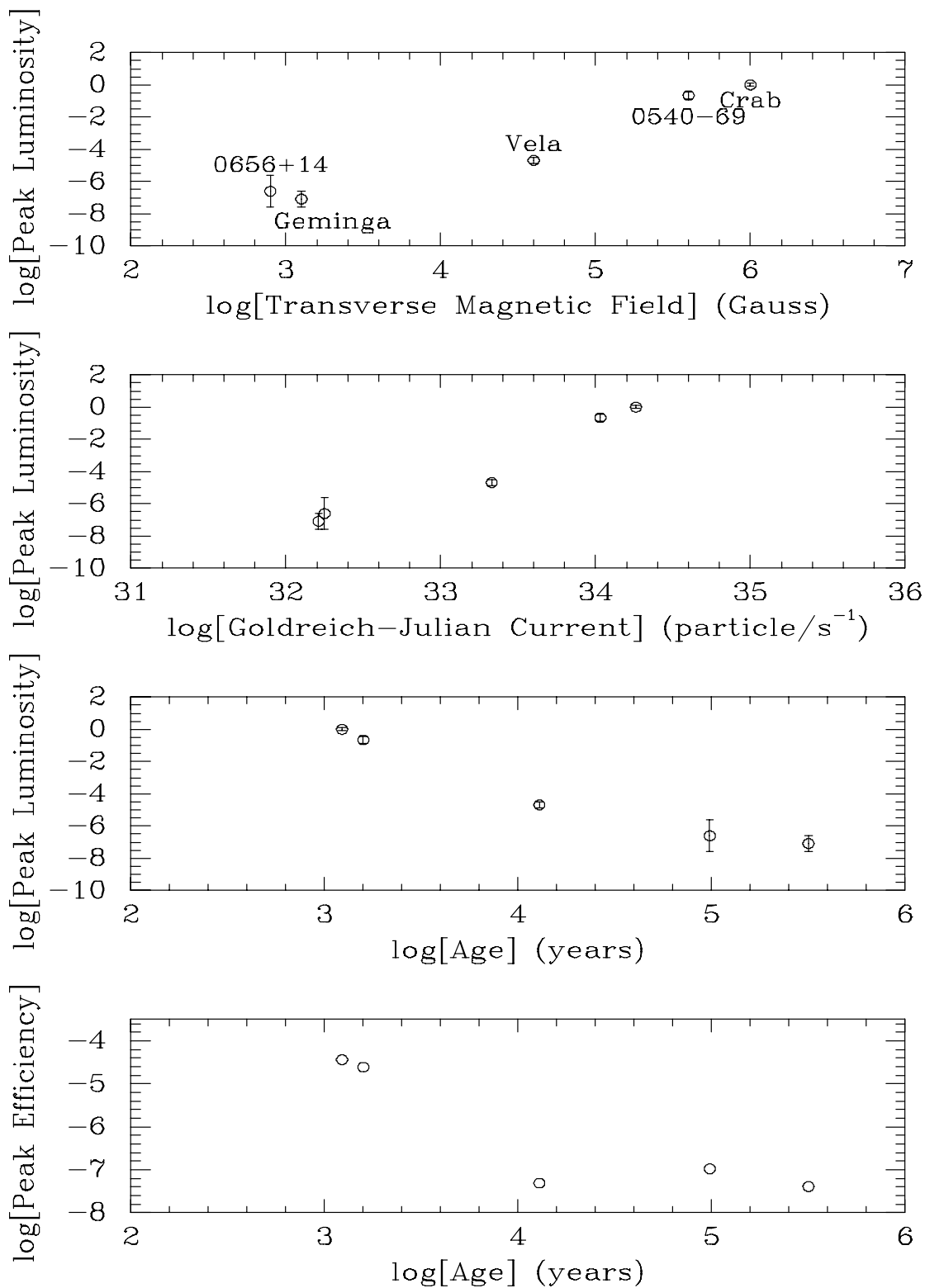} \caption{Peak optical luminosity
vs light cylinder field, Goldreich \& Julian current and canonical
age. Also shown is the efficiency of the peak emission against age.
The peak luminosity has been normalised to the Crab pulsar. The error
bars represent both statistical errors from the pulse shape and
uncertainty in the pulsar distance.} \end{figure}

\begin{figure*}
\plotone{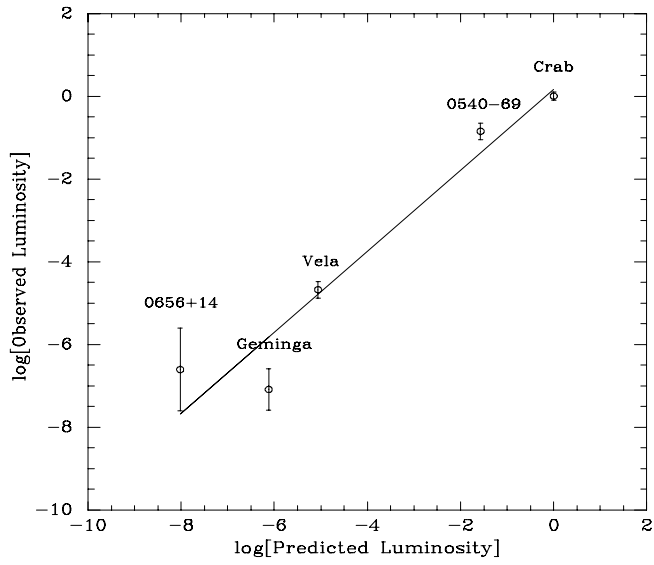}
\caption{Predicted peak optical luminosity from equation 2 versus
observed peak emission. Also shown is the weighted fit described in the
text. The weighting was based upon the observational uncertainities
in flux and distance.}
\end{figure*}

\begin{figure*}
\plotone{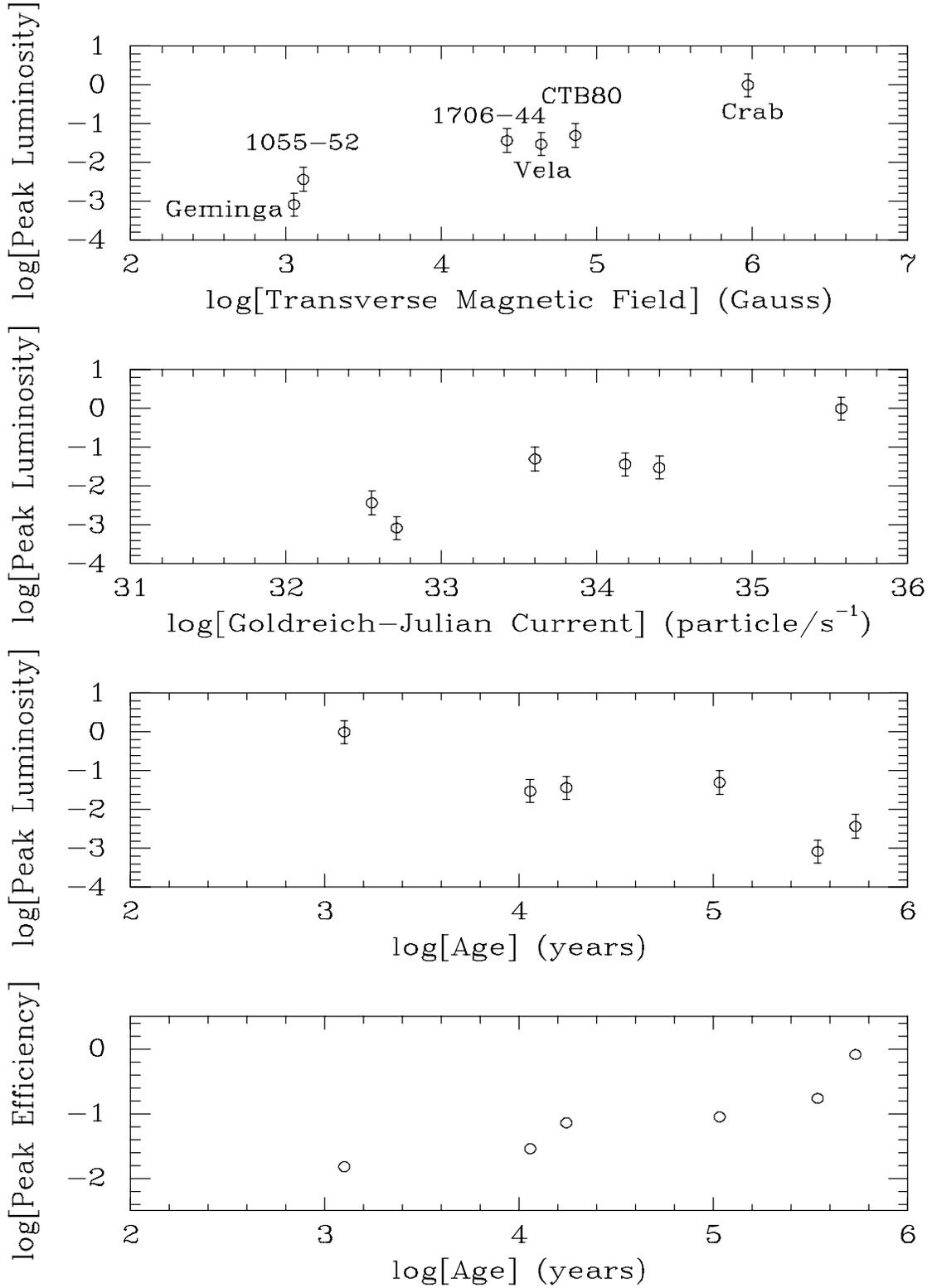}
\caption{ Peak $\gamma$-ray luminosity vs light cylinder field, Goldreich
\& Julian current and canonical age. Also shown is the efficiency of
the peak emission against age. The peak luminosity has been normalised
to the Crab pulsar. The $\gamma$-ray peak luminosity has been inferred
from \cite{fie98}, \cite{thom96} and \cite{thom99}.}
\end{figure*}

\begin{onecolumn}
\begin{table}

\caption{Main characteristics of Optical Pulsars: B$_S$ \& B$_{LC}$,
the canonical surface and transverse magnetic field at the light
cylinder respectively; Int and Peak refer to the optical luminosity,
integrated and peak, at the indicated distance in the B band. }

\begin{tiny}

\begin{tabular}{lcccccccccc}
Name & D      & P  & $\dot{P}$  & Log Age & $B_S$ & $B_{LC}$  & Int & Peak & Spec. Index & Cutoff \\
     &  (kpc) & (ms)   &  $10^{-14}$ s/s & Years & log(G) & log (G)   & $\mu$Crab & $\mu$Crab & at 4500 \AA & \AA \\

Crab        & 2       & 33  & 42 & 3.09  & 12.6 & 6.1  & $10^6$ & $10^6$ & -0.11 & 15000(?)\\
Vela        & 0.5     & 89  & 11  & 4.11 & 12.5 & 4.8  & 27 & 21  & 0.2 & 6500(?) \\
PSR0545-69  & 49      & 50  & 40  & 3.20 & 12.7 & 5.7  & $1.1~10^6$ & $1.4~10^5$  & 0.2 & $>$7000 \\
PSR0656+14  & 0.25(?) & 385 & 1.2 & 5.50 & 12.7 & 3.0  & 1.8 & 0.3  & 1.3 & $>$8000 \\
PSR0633+17     & 0.16    & 237 & 1.2 & 4.99 & 12.2 & 3.2  & 0.3 & 0.1  & 1.9 & $>$8000 \\
\end{tabular}
\end{tiny}
\end{table}
\end{onecolumn}

\begin{onecolumn}
\begin{table}

\caption{Main characteristics of $\gamma$-ray Pulsars: B$_S$ \&
B$_{LC}$ the canonical surface and transverse magnetic field at the
light cylinder respectively; $\gamma$ Int and Peak refer to the
integrated and peak $\gamma$-ray luminosity at the indicated distance
for E $>$ 100 MeV. }

\begin{tiny}

\begin{tabular}{lcccccccc}
Name & D      & P  & $\dot{P}$  &  $B_S$ & $B_{LC}$  & Int Lumin. & Peak Lumin. & Spectral Index  \\
     &  (kpc) & (ms)   &  $10^{-14}$ s/s & log(G) & log (G)   & Crab=1 & Crab=1 & \\

Crab           & 2       & 33  & 42.1 & 12.58 & 5.97  & 1 & 1 & 2.15 \\
Vela           & 0.5     & 89  & 12.5  & 12.53 & 4.64  & 0.0480 & 0.0299  &  1.70 \\
PSR1055-52     & 0.5-1.5 & 197  & 0.6  & 12.03 & 3.11  & 0.0124 & 0.0037  & 1.18 \\
PSR1706-44     & 2.4     & 102 & 9.3 & 12.49 & 4.42  & 0.1380 & 0.0368  & 1.72 \\
PSR0633+17     & 0.16    & 237 & 1.1 & 12.21 & 3.05  & 0.0019 & 0.0008  & 1.50 \\
PSR1951+32     & 2.5     & 40 & 0.6 & 11.69 & 4.86  & 0.0500 & 0.0187  & 1.74 \\
\end{tabular}
\end{tiny}

\end{table}
\end{onecolumn}

\begin{onecolumn}
\begin{table}

\caption{Predicted Luminosity of X and $\gamma$-ray emitting
pulsars. The duty cycle has been estimated from $\gamma$-ray
observations. Also included are the nearby millisecond pulsar PSR
J2322+2057 (\cite{nic93}) and the anomolous X-ray pulsar 1E1841-045 (\cite{vas97}).}

\begin{tiny}

\begin{tabular}{lccccccc}
Name          & D      & P      & $\dot{P}  $     &  $B_S$ & $B_{LC}$   &  Duty & Pred. Lumin. \\
              &  (kpc) & (ms)   &  $10^{-14}$ s/s & log(G) & log (G)   & Cycle &  $\mu$Crab  \\

PSR1055-52    & 1.5    & 197    & 0.6             & 12.03  & 3.11      & 0.2    &  0.01 \\
PSR1706-44    & 1.8    & 102    & 9.3             & 12.49  & 4.42      & 0.14    & 35  \\
PSR1951+32    & 2.5    & 40     & 0.6             & 11.69  & 4.86      & 0.08    & 670    \\
PSR1821-24(M28) &  5.1   & 3      & 1.1$^{-4}$      & 9.3  & 5.8      & 0.1      & 0.3-1 10$^6$  \\
PSRJ2322+2057  &  0.78  & 4.8    & 7.0$^{-7}$      & 8.3  & 4.2      & 0.3(?)      & 8 \\
1E1841-045    & 7      & 11770  & 4700            & $\approx 15$  & -2      & 0.5?     & $<<10^{-10}$  \\

\end{tabular}
\end{tiny}

\end{table}
\end{onecolumn}


\begin{thebibliography}{}
\bibitem[Becker \& Tr\"umper (1997)]{bec97} Becker, W. \& Tr\"umper, J., 1997, \aap, 326, 682 
\bibitem[Bignami et al (1988)]{big88} Bignami, G. F., Caraveo, P. A. \& Paul, J. A., 1988, \aap, 202, L1 
\bibitem[Bignami et al (1996)]{big96}Bignami, G. F., Caraveo, P. A., Mignani, R., Edelstein, J. \& Bowyer, S, 1996, \apj, 456, L111 
\bibitem[Cocke et al (1969)]{coc69}Cocke, W. J., Disney, M. J. \& Taylor, D. J., 1969, Nature, 221, 525 
\bibitem[Caraveo et al (1996)]{car96}Caraveo, P. A., Bignami, G. F., Mignani, R. \& Taff, L. G., 1996, \aaps, 120, 65 
\bibitem[Eikenberry \& Fazio (1997)]{eik97}Eikenberry, S. S. \& Fazio, G. G., 1997, \apj, 476, 281 
\bibitem[Fierro et al (1998)]{fie98}Fierro, J. M., Michelson, P. F., Nolan, D. C. \& Thompson, D. J., 1998, \apj, 494, 734
\bibitem[Golden \& Shearer (1999)]{gold99}Golden, A. \& Shearer, A., 1999, \aap, 342, L5 
\bibitem[Golden et al (2000)]{gold00}Golden, A., Shearer, A., 2000, \apj, 999, 99 
\bibitem[Golden et al (2000)]{gold00a}Golden, A., Shearer, A. \&Redfern, M, 2000, submitted to \aap 
\bibitem[Goldoni et al (1995)]{gold95}Goldoni, P., Musso, C., Caraveo, P. A. \& Bignami, G. F., 1995, \aap, 298, 535
\bibitem[Goldreich \& Julian (1969)]{gold69}Goldreich, P. \& Julian, W., 1969, \apj, 245, 267
\bibitem[Gouiffes \& \"{O}gelman(2000)]{gou00}Gouiffes, C. \& \"{O}gelman, H, 2000, Pulsar Astronomy - 2000 and Beyond, ASP Conference Series, Vol. 202; Proceedings of the 177th IAU  Colloquium 177 (San Francisco: ASP). Edited by M. Kramer, N. Wex, and N. Wielebinski, p. 301 
\bibitem[Jacchia et al (1999)]{jac99}Jacchia, A., de Luca, F., Lazzaro, E., Caraveo, P. A., Mignani, R. P. \& Bignami, G. F., 1999, \aap, 347, 494
\bibitem[Hulleman et al (2000)]{hull0}Hulleman, F., van Kerkwijk, M. H., Verbunt, F. W. M., Kulkarni, S. R., 2000,\aap, 358, 606
\bibitem[Halpern \& Tytler (1988)]{hal88} Halpern, J-P \& Tytler, D., 1988, \apj, 330, 201 
\bibitem[Marshall et al (1998)]{mars98} Marshall, F. E., Gotthelf, E. V., Zhang, W., Middleditch, J. \& Wang, Q. D., 1998, \apj, 499, L179 
\bibitem[Martin et al (1998)]{mar98} Martin, C. Halpern, J. P. \& Schiminovich, D, 1998, \apj, 494, 211
\bibitem[McLaughlin \& Cordes (2000)]{mcl00}McLaughlin M. A. \& Cordes, J. M., 2000, \apj, 538, 818
\bibitem[Middleditch \& Pennypacker (1985)]{mid85}Middleditch, J. \& Pennypacker, C., 1985, Nature, 313, 659 
\bibitem[Mignani et al (1998)]{mig98}Mignani, R. P., Caraveo, P. A., \& Bignami, G. F., 1998, \aap, 332, L37
\bibitem[Mignani et al (2000)]{mig00}Mignani, R. P., Pulone, L., Marconi, G. Iannicola, G. \& Caraveo, P. A., 2000, \aap, 355, 603 
\bibitem[Nasuti et al (1997)]{nas97}Nasuti, F. P., Mignani, R., Caraveo, P. A. \& Bignami, G. F., 1997, \aap, 323, 839
\bibitem[Nice et al (1993)]{nic93} Nice, D. J., Taylor, J. H. \& Fruchter, A. S., 1993, \apj, 402, L49
\bibitem[Pavlov et al (1997)]{pav97}Pavlov G. G., Welty, A. D. \& Cordova, F. A., 1997, \apj, 489, L75
\bibitem[Pacini (1971)]{pac71}Pacini, F., 1971, \apj, 163,17 
\bibitem[Pacini \& Salvati (1983)]{pac83}Pacini, F.  \& Salvati, M., 1983, \apj, 274, 369 
\bibitem[Pacini \& Salvati (1987)]{pac87}Pacini, F.  \& Salvati, M., 1987, \apj, 321, 445 
\bibitem[Perryman et al (1999)]{per99}Perryman, M. A. C., Favata, F., Peacock, A., Rando, N. \& Taylor, B. G., 1999, \aap, 346, 30 
\bibitem[Romani \& Yadigaroglu (1995)]{rom95}Romani, R. W., \& Yadigaroglu, I.-A., 1995, \apj, 438, 314  
\bibitem[Romani et al (1999)]{rom99}Romani, R. W., Miller, A. J., Cabera, B. \& Figueroa-Feliciano, E., 1999, \apj, 521, L151 
\bibitem[Shearer et al (1997)]{shear97}Shearer, A., Redfern, R.  M., Gorman, G., Butler, Golden, A., R.,
 O'Kane, P., Golden, A., Beskin, G.  M., Neizvestny, S.  I.,
 Neustroev, V.  V., Plokhotnichenko, V.  L.  \& Cullum, M., \apj, 1997, 487, L181 
\bibitem[Shearer et al (1998)]{shear98}Shearer, A., Harfst, S., Redfern, R.  M., Butler, R., O'Kane, P.,
 Beskin, G.  M., Neizvestny, S.  I., Neustroev, V.  V.,
 Plokhotnichenko, V.  L.  \& Cullum, M., 1998, \aap, 335, L21 
\bibitem[Smith et al (1988)]{smith88}Smith, F. G., Jones, D. H. P., Dick, J. S. P. \& Pike, C. D., 1988, \mnras, 233, 305 
\bibitem[Thompson et al (1996)]{thom96}Thompson, D. J. et al, 1996, \apj, 465, 385
\bibitem[Thompson et al (1999)]{thom99}Thompson, D. J. et al, 1999, \apj, 516, 297
\bibitem[Vasisht \& Gotthelf]{vas97}, Vasisht, G. \& Gotthelf, E. V., 1997, \apj, 486, L129
\bibitem[Wallace et al (1977)]{wall77}Wallace, P. T. et al. 1977, Nature, 266, 692 
\bibitem[Wei, Song \& Lu (1997)]{wei97} Wei, D. M., Song, L. M. \& Lu, T., 1997,\aa, 323, 98
\end{thebibliography}
\end{document}